\documentstyle[12pt,a41,epsfig,wrapfig]{article}

\begin{document}
\begin{flushright}
DESY 98-032\\
hep-ph/9803447
\end{flushright}
\vspace*{3cm}

\begin{center}
{\Large \bf Color Octet Contribution to $J/\psi$ Production at a Photon
  Linear Collider} \\
\vspace*{2cm}
{G.~JAPARIDZE$^a$, A. TKABLADZE$^{b}$\footnote{Alexander von Humboldt
    Fellow; e-mail: avto@ifh.de.} }\\
\vspace*{1.cm}
{\it $^a$Center for Theoretical studies of Physical
systems, \\ Clark
Atlanta University, Atlanta, GA 30314, U.S.}\\
\vspace*{0.5cm}
{\it $^b$DESY, D-15738, Zeuthen,  Germany}\\
{\it and}\\
{\it Bogoliubov Laboratory of Theoretical Physics,}\\
{\it JINR, Dubna, Russia}

\vspace*{3.5cm}

\end{center}

\begin{abstract}
{\small    
We investigate $J/\psi$ production at a Photon Linear Collider (PLC)  in the
framework of NRQCD factorization approach.
 It was shown that color octet contribution is dominating
in charmonium production at PLC. The main fraction of $J/\psi$ comes from the
single and double resolved photon processes. Expected cross sections of
charmonium states production are large enough even at $p_T>5$ GeV to
expect several hundreds of  $J/\psi$'s through the leptonic decays channel for
a projected integral PLC luminosity of $50 fb^{-1}$}
\end{abstract}

\newpage

\section{Introduction}
\setcounter{equation}{0}
\vspace{-3mm}

The nonrelativistic QCD (NRQCD) factorization approach, developed in
the last few years,  is the 
most successful method for the analysis of  heavy quarkonium production
and decay processes \cite{NRQCD}.
In the framework of this formalism cross section and decay rate of the 
process involving heavy quarkonium is
factorized into the short distance part (coefficient functions) and the long distance 
one. For the cross section $\sigma(A+B\to H)$, where $H$ stands for
quarkonium, the formalism states:
\begin{eqnarray}
\sigma(A+B\to H) = \sum_{n}{\sigma(A+B\to Q\bar Q[n])
\langle0|{\cal O}^{H}[n]|0\rangle}.
\end{eqnarray}
 In (1) $\sigma(A+B\to Q\bar Q[n])$ is the cross section of the production
of the heavy quark-antiquark pair state $[n]$. 
This part of cross section can be calculated perturbatively.
 The  matrix element  $\langle0|{\cal O}^H[n]|0\rangle$
stands for the evolution of the state $[n]$ into the 
 particular hadronic state $H$.
It is defined by long distance dynamics and cannot
be computed perturbatively. The long distance matrix elements scale as
a power of $v$ - the  relative velocity of 
the heavy constituents in the quarkonium. 
So, in the NRQCD factorization approach (FA) the cross section of  heavy
quarkonium production is considered as a double series in the QCD
coupling constant and the relative velocity $v$.
 The relative importance of long distance matrix
elements in powers of $v$ can be estimated  using
the NRQCD velocity scaling rules \cite{LMNMH}.

In the NRQCD factorization approach the  quark-antiquark color 
octet intermediate  states are allowed to contribute 
in heavy quarkonium  production and decay processes  in 
higher order. By such a way the Color Octet Mechanism (COM)  takes into
account  the complete structure of the quarkonium Fock space while in 
the color singlet model (CSM) only the dominant Fock state is 
considered \cite{CSM}.
The shape of the  $p_T$ distribution of the $^3S_1^{(8)}$ octet state
production cross section indicates that $J/\psi$ and $\psi'$
production at large $p_T$ observed at the Tevatron (FNAL)
can be explained in the framework of FA \cite{BF,CL}.  

Despite the obvious successes of the COM  some problems 
remain unsolved.
 In particular, the theoretical
predictions disagree  with  the $J/\psi$ and $\psi'$ polarization data at
fixed target energies \cite{BeR,Beneke} and the COM estimation for the
yield ratio of $\chi_{c1}$ and $\chi_{c2}$ states remains too low \cite{BeR}.
These discrepances
indicate that higher twist corrections might contribute significantly in 
production of charmonium states at low $p_T$ and should be considered 
equally with the COM contributions \cite{VHBT}.
The latter underestimate the $J/\psi$
photoproduction cross section at large values of $z$ in  H1 and ZEUS
experiments at HERA
($z=E_{J/\psi}/E_{\gamma}$ in the proton rest frame)- 
the most problematic issue of the COM \cite{CK}.

The NRQCD factorization approach has predictive power when the
relative velocity $v$ is small enough, allowing to describe with reasonable
accuracy several experimental data by taking into account only a few lower
order terms in powers of $v$.   
Hence it becomes necessary to check the  universality of the FA,  stating 
that the values
of long distance matrix elements, extracted from the 
experimental data  for different processes, should be the same.
Unfortunately, due to the 
rather large theoretical uncertainties at the present time, the
existing  experimental data does not allow  to extract the values of 
the long distance matrix elements with  enough accuracy.

This has motivated us to look for  processes with less
theoretical uncertainties to test the color octet mechanism.
The measurements  of  the $J/\psi$ polarization
in  unpolarized hadron-hadron collisions and electroproduction 
\cite{VHBT,BK, FleN,TT} can be used for these purposes
 as well as the asymmetries in $J/\psi$ production
\cite{TT}. 
Unlike  hadron-hadron collisions,  existing $J/\psi$
photoproduction data can be explained within the CSM taking into
account higher order QCD corrections \cite{KraCSM}. Moreover, color
octet contributions overestimate the experimental data drastically
\cite{CK}. In this sense, the study of $J/\psi$ production 
processes in non-hadron collisions can clarify the existing
contradictory picture. 

The collision of high-energy, high-intensity photon beams at a
photon linear collider, obtained via Compton backscattering of 
laser beams off linac electron beams, should provide another possibility
to investigate the heavy quark-antiquark bound states production processes.
Based on a $e^+e^-$ linear collider, the PLC will have almost the
same energy and a luminosity, i.e. a c.m. energy of 500-1000 GeV and
a luminosity of the order of $10^{33} cm^{-2} s^{-1}$ \cite{11JT}.

In the present paper we study $J/\psi$ production at a photon linear
collider. The color octet contribution in  $J/\psi$ production at PLC was
calculated recently \cite{MMP}, where only the direct (unresolved)
photon-photon collisions were considered. We will show that the resolved
photon processes are more significant because there exist diagrams in
lowest order of perturbative QCD leading to the 
subprocess cross section, non decreasing when the invariant mass of
photon-photon system increases. 
In the next section we present  all possible subprocesses in  lowest 
order in perturbative QCD and fix the color octet long distance 
matrix elements needed.
 The numerical results are given in Sec. III. It is  shown the
that color octet contribution in resolved photon subprocesses is essential
and the expected  cross section of charmonium production is large
enough even at high $p_T$ ($>5$ GeV) allowing to extract valuable information
about color octet long distance matrix elements.

\section{$J/\psi$ Production Subprocesses and COM Parameters}
\vspace{-3mm}

For the calculation of the $J/\psi$ production cross section at PLC we
consider the following subprocesses: direct $\gamma\gamma$ collision,
single  and double resolved photon processes, when the 
 heavy quark-antiquark pair is produced
via hadronic components of the photon, which are described in terms of the
photon structure functions \cite{Witten}.

To lowest order in $\alpha_{em}$ and $\alpha_S$ the following subprocesses 
contribute:
\begin{eqnarray}
\gamma+\gamma \to (c\bar c) + \gamma (g) \\
\gamma+\gamma(g,q) \to (c\bar c)+g,q \\
\gamma(g,q)+\gamma(g,q) \to (c\bar c)+g,q  
\end{eqnarray}
where $\gamma(g,q)$ denotes a gluon or quark component of the photon,
$(c\bar c)$ stands for  any possible quark-antiquark state.

The cross section of $J/\psi$ production in the NRQCD factorization scheme
can be presented as

\begin{eqnarray}
\sigma_{J/\psi} = \sigma(J/\psi)_{dir} +
\sum_{J=0,1,2}{Br(\chi_{cJ}\to J/\psi X)\sigma_{\chi_{cJ}}}+Br(\psi'\to J/\psi X)
\sigma_{\psi}
\end{eqnarray}
Here $Br((c\bar c)\to J/\psi X)$ denotes the branching ratio of the
corresponding $(c\bar c)$ state into $J/\psi$.
To the production of each state of quarkonium both color
octet and color singlet states contribute, as in the case of direct $J/\psi$
production
\begin{eqnarray}
\sigma(J/\psi)_{dir} = \sigma_{J/\psi}^{singl}+\sigma^8_{J/\psi}=
\sigma(J/\psi)^{singl}+
\sum{\sigma(Q\bar Q[^{2s+1}L_J^{8}])
\langle0|{\cal O}_8^{J/\psi}(^{2s+1}L_J)|0\rangle}
\end{eqnarray}
where the sum runs over the states $^3P_{0,1,2}^8$, $^1S_0^8$ and $^3S_1^8$.
In the framework of the NRQCD FA velocity expansion we consider only the 
contribution of the dominant sets of color octet states in the direct 
production of $S$ and $P$ states of charmonium.

In direct $\gamma\gamma$ interaction
in  lowest order only the two states of heavy quark-antiquark pairs can
be produced :  $^3S_1$ color
singlet ($J/\psi$) and $^3S_1$ color octet \cite{MMP}.
For the differential cross section of the singlet state production we
have:
\begin{eqnarray}
\frac{d\sigma_{\gamma\gamma}}{dt} = && 
\frac{16 M_{J/\psi}}{3} \bigl(\frac{2}{3}\bigr)^4 (4\pi)^2
\alpha_S\alpha^2
\langle{\cal O}_1^{J/\psi}(^3S_1)\rangle \nonumber \\
&&\frac{\hat s^2(\hat s-M_{J/\psi}^2)^2+
      \hat t^2(\hat t-M_{J/\psi}^2)^2  +
      \hat u^2(\hat u-M_{J/\psi}^2)^2}
{\hat s^2(\hat s-M_{J/\psi}^2)^2 (\hat t-M_{J/\psi}^2)^2
(\hat t-M_{J/\psi}^2)^2}
\label{singlet}
\end{eqnarray}
where $\hat s$, $\hat t$ and $\hat u$ are the standard Mandelstam
variables. 

The differential cross section for the $J/\psi$ production through
color octet $^3S_1$ state can be obtained from  (\ref{singlet}) by the
replacement \footnote{The additional factor  4/3 coming from different
  color factors of considered processes is omited in the paper \cite{MMP}.}
\begin{eqnarray}
\langle{\cal O}_8^{J/\psi}(^3S_1)\rangle \to
\frac{4}{3}\frac{4 \alpha}{9 \alpha_S}
\langle{\cal O}_1^{J/\psi}(^3S_1)\rangle 
\nonumber
\end{eqnarray}

We calculated all possible subprocesses for single resolved photon reaction.
Our results are in agreement with \cite{CK}.
The subprocess  for $(c\bar c)$  states
production in double resolved photon case are same as for 
$J/\psi$ production in hadron-hadron collision and were calculated by
Cho and Leibovich \cite{CL}.


All  color singlet long distance matrix elements are related to the radial quarkonium wave
functions at the origin and their derivatives.We have 
used  the Buchm\"uller-Tye wave functions
at the origin, tabulated in \cite{BeR}.
 The number of color octet long distance matrix elements can be reduced by
 using the NRQCD spin symmetry relations:
\begin{eqnarray}
 \langle0|{\cal O}_8^H(^3P_J)|0\rangle &=& (2J+1)\langle0|{\cal O}_8^H(^3P_0)|0\rangle,\\
 \langle0|{\cal O}_8^{\chi_{cJ}}(^3S_1)|0\rangle &=& (2J+1)\langle0|{\cal O}_8^{\chi_{c0}}(^3S_1)|0\rangle.
\end{eqnarray}
These relations are valid up to order $v^2$.

After using these relations only three independent matrix elements remain - 
$\langle{\cal O}_8^{J/\psi}(^3S_1)\rangle$,
$\langle{\cal O}_8^{J/\psi}(^3P_0)\rangle$ 
and $\langle{\cal O}_8^{J/\psi}(^1S_0)\rangle$, which give the main
contribution to the direct $J/\psi$ production cross section.
The values of these parameters are extracted from the data of direct $J/\psi$
production at CDF \cite{CL,BK}. We used the
results of paper \cite{BK}, where a more detailed numerical analysis was
carried out:
\begin{eqnarray}
 &&  \langle0|{\cal O}^{J/\psi}_8(^3S_1)|0\rangle= 1.06\pm0.14^{+1.05}_{-0.59}\cdot10^{-2}
    GeV^3,\\
&&\langle0|{\cal O}_8^{J/\psi}(^1S_0)|0\rangle+\frac{3.5}{m_c^2}\langle0|{\cal O}_8^{J/\psi}(^3P_0)|0\rangle =
3.9\pm1.15^{+1.46}_{-1.07}\cdot10^{-2} GeV^3
\end{eqnarray}
Further we will assume that 
$\langle0|{\cal O}_8^{J/\psi}(^1S_0)|0\rangle=3.5/m_c^2\langle0|{\cal
  O}_8^{J/\psi}(^3P_0)|0\rangle$. 
For the indirect $J/\psi$ production via the  $\chi_{cJ}$ states
decays we use the following value fitted from the CDF data \cite{CL}:
\begin{eqnarray}
\langle{\cal O}_8^{\chi_{c1}}(^3S_1)\rangle &= &9.8\cdot10^{-3} GeV^3.
\end{eqnarray}

We would like to to mention that all these values contain
 large uncertainties (for more detailed
analyses see \cite{CL,BK,TrigB}) and  we use them only to estimate the expected
cross sections of the production of different charmonium states  at PLC.

\section{Numerical Results}

At PLC the energy distributions of the colliding photon beams are
not monochromatic. One of the main characteristics of the Compton
scattering process is the dimensionless variable $x$,
$$
 x = \frac{4 E_b \omega_0}{m_e^2} \cos{\theta/2}.
$$
Here $E_b$ is the linac beam energy, $\omega_0$ is the energy of the
laser photons and $\theta$ is the angle between the directions
 of the electron and laser beams.
The maximum  energy of the obtained photon beam depends on  $x$,

$$
\omega_{max} = E_b \frac{x}{1+x},
$$ and increases with  increasing $x$. 
The value $x=4.83$ is optimum for PLC to prevent $e^+e^-$ pair
creation in a collision of laser and backscattered photon beams
\cite{Telnov}. In our calculations we use  $x=4.8$. This means that the
maximum energy fractio carried by backscattered photon, $\omega_b/E_b$,
is about 0.83.

The luminosity distribution of the photon beam depends
sensitively on the conversion distance due to non-zero angles of 
photon-electron beam scattering.
The conversion distance is the
distance from the laser  and electron beam intersection point to the
final interaction point. The luminosity distribution depends also on
the electron beam size and shape.
Assuming round Gaussian linac beams, the luminosity spectrum is
characterized by a geometrical factor $\rho$, the ratio of the intrinsic
transverse spread of the photon beam to that of the original one :
\begin{eqnarray}
\rho = \frac{l\theta_0}{\sqrt{2}\sigma_e} = 
3.61\sqrt{x+1}
\biggl(\frac{l}{cm}\biggr)\biggl(\frac{E_b}{TeV}\biggr)^{-1}
\biggl(\frac{\sigma_e}{nm}\biggr)^{-1},
\end{eqnarray}
\begin{wrapfigure}{l}{9.cm}
\vspace*{-7mm}
\centering
\epsfig{file=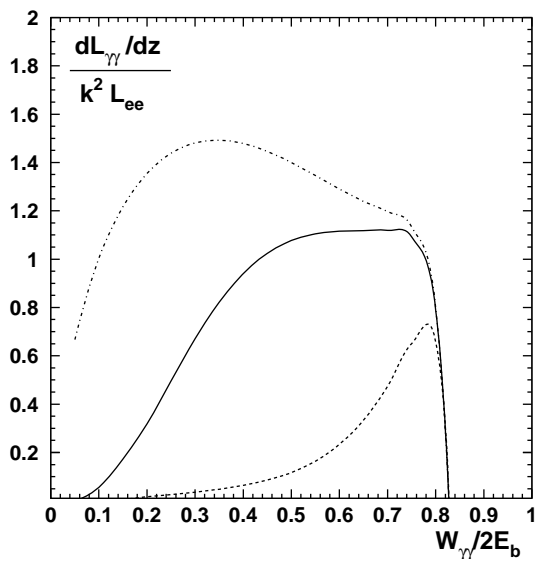,width=8.cm}
\caption{\small
Spectral luminosity of $\gamma\gamma$ collisions; the dashed line
corresponds to the value $\rho=0$, the solid line to $\rho=0.6$ and
the dashed-dotted line $\rho=2.4$.
}
\end{wrapfigure}
 here $l$ is the conversion distance and $\theta_0$ is defined by the
 electron beam energy, $\theta_0= m_e\sqrt{1+x}/E_b$.
In fig.1 the  spectral luminosities for unpolarized 
photon and electron beams are presented for 
different values of $\rho$ are presented.
The parameter $k$  is the ratio of the number of electrons
scattered  on the laser beam to the number of electrons in  
each electron bunch .  
With $\rho$ increasing the spectral density becomes
narrower and more energetic photons dominate, meanwhile the total photon
photon luminosity is decreasing. 
To achieve a maximum  $J/\psi$ production cross section 
the geometrical
factor $\rho$ should be as small as possible. However, in this case 
the transverse separation of the electron and the secondary photon beams 
becomes impossible and 
large background is expected from the process 
$e\gamma\to eZ\to eb\bar b$ with the subsequent decays of $b(\bar b)$
quarks into $J/\psi$. 
Hence we use $\rho=0.6$  which is small
 enough to achieve an observable $J/\psi$ production cross section (see
 below) as well as large enough to eliminate electron-photon collision
 background \cite{BBC}.

\vspace{2mm}
{\bf Table 1}. $J/\psi$ production cross sections in direct
$\gamma\gamma$
interaction at PLC for $\sqrt{s_{e^+e^-}}=500$ GeV and $800$ GeV.
The cross sections are calculated for two different values of the
geometrical factor $\rho=$0 and 0.6.

\begin{center}
\begin{tabular}{|c|c|c|c|c|}
\hline \hline
$\rho$ & \multicolumn{2}{c}{$\sqrt{s}$=500 GeV}\vline&
\multicolumn{2}{c}{$\sqrt{s}$=800 GeV}\vline\\
\cline{2-5} 

 & singlet & octet & singlet &octet \\
\hline
 0.  & $2.79\cdot10^{-5} nb $ & $2.71\cdot10^{-5} nb$
 & $1.21\cdot10^{-5} nb$ & $1.18\cdot10^{-5} nb$\\
\hline 
 0.6 & $1.74\cdot10^{-7} nb$ & $1.69\cdot10^{-7} nb$
 & $3.10\cdot10^{-8} nb $& $3.00\cdot10^{-8} nb$\\
\hline 
\end{tabular}
\end{center}

In Table I $J/\psi$ production cross sections for two
different values of the geometric factor, $\rho=0$ and $0.6$, are presented.
 We have used the color octet long
distance parameters presented in the previous section.
As one can see from table I, the total cross section of $J/\psi$
production is decreasing with increasing  $e^+e^-$ collision energy. 
The reason for this is that
the  production of the $^3S_1$ state (singlet or
octet ) in direct $\gamma\gamma$ interaction falls off as
$1/s^2$ with increasing  photon-photon invariant mass.
 For  $\rho=0$ the luminosity of low
energetic photons is higher than for large values of $\rho$ and 
correspondingly the $J/\psi$ production cross section is higher.
The $J/\psi$ production cross section in direct photon-photon collision
for $\rho=0.6$ is practically unobservable even at a high projected
$e^+e^-$ collider luminosity of $100 fb^{-1}$.

As for the resolved photon subprocess, unlike direct
$\gamma\gamma$
collisions, there exist diagrams with only vector particles (gluon) in the
$t$-channel in the lowest order of perturbative QCD. These diagrams
result in a cross section of $(c\bar c)$ pair
production which is not decreasing with increasing $s_{\gamma\gamma}$.

\begin{figure}[ht]
\vspace{-4mm}
\centering
\begin{minipage}[c]{7.5cm}
\centering
\epsfig{file=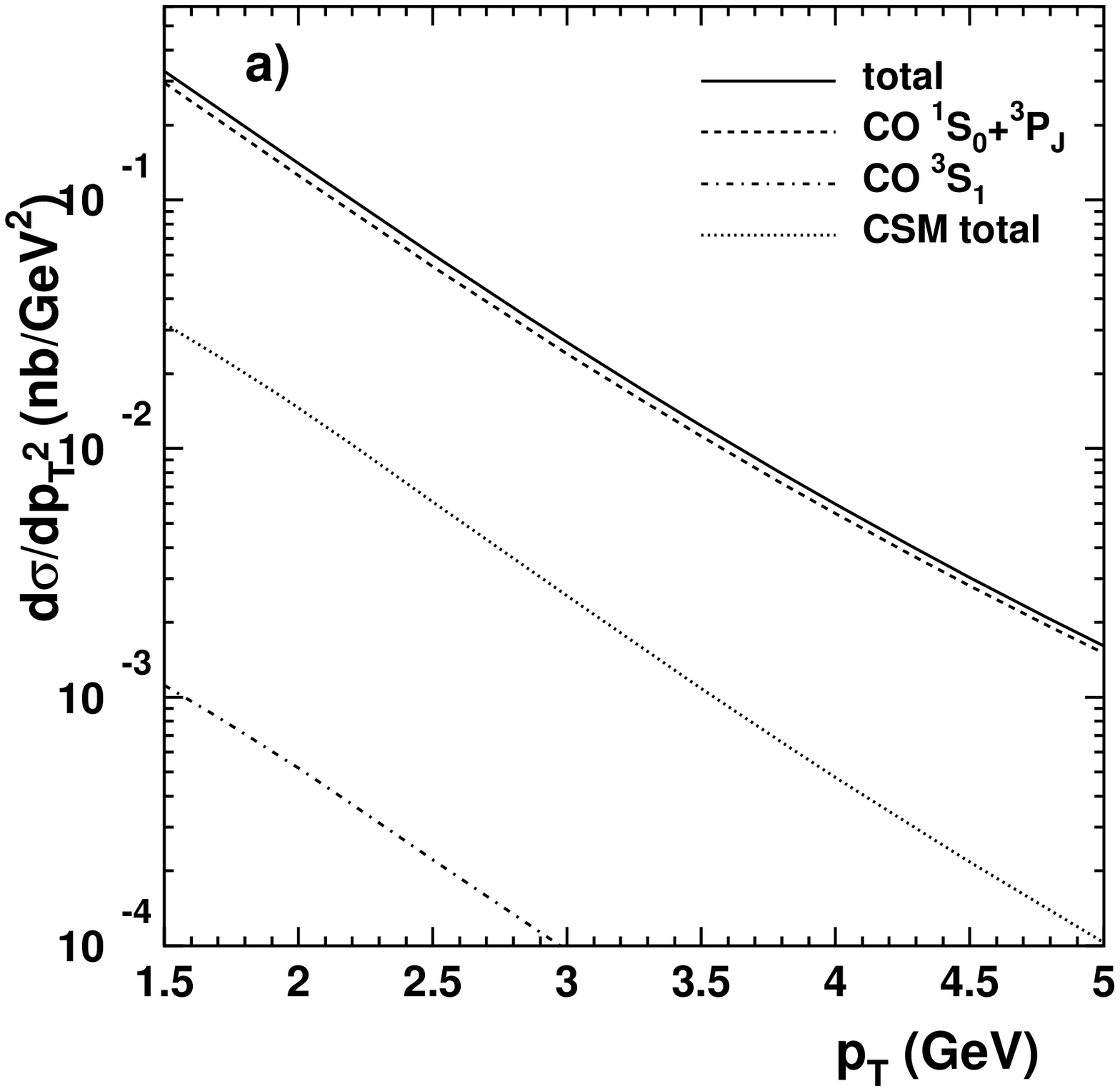,width=7.5cm}
\end{minipage}
\hspace*{0.5cm}
\begin{minipage}[c]{7.5cm}
\centering
\epsfig{file=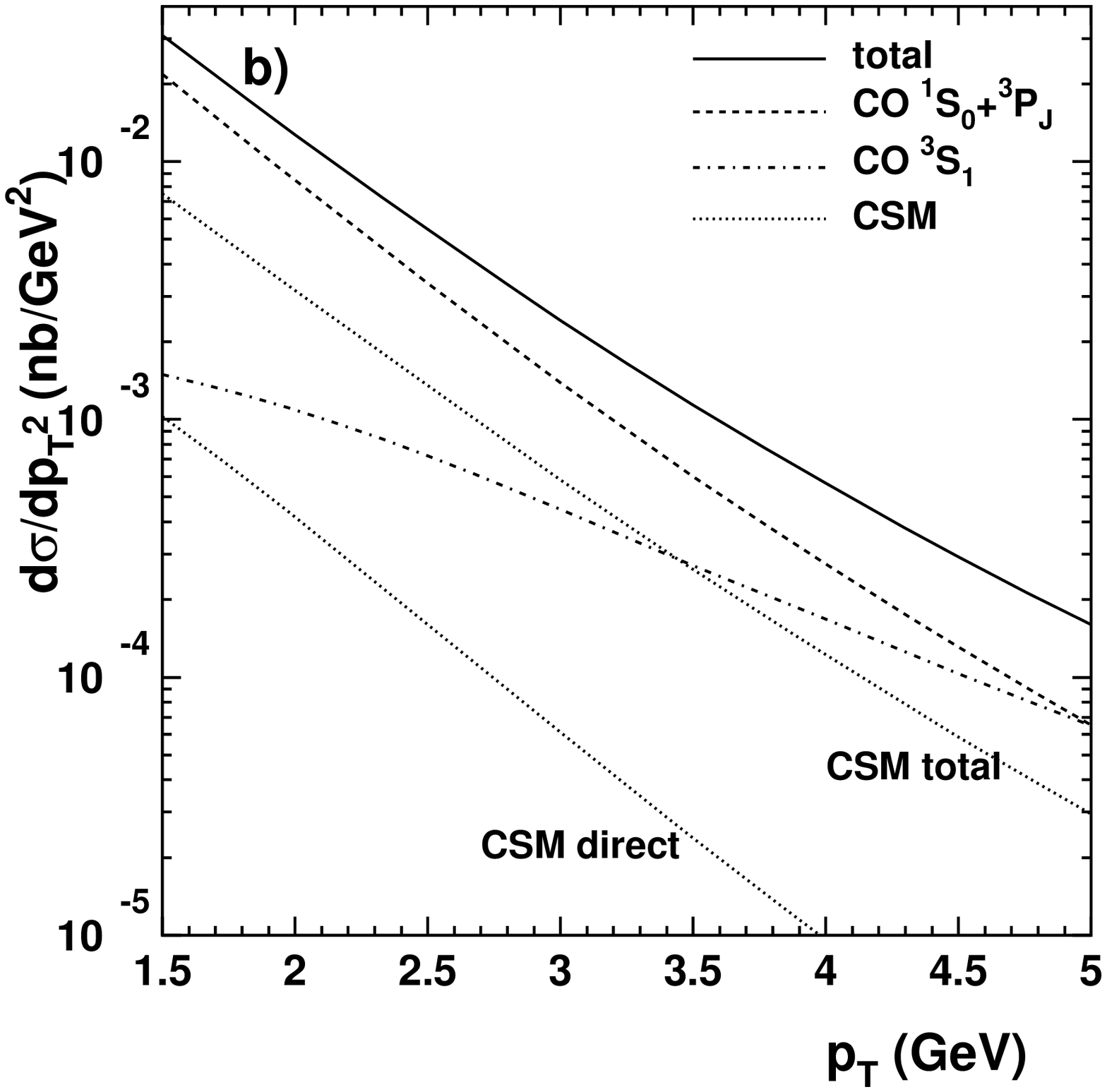,width=7.5cm}
\end{minipage}
\caption{\small
The $J/\psi$ production cross sections at $\sqrt{s_{e^+e^-}}=500$ GeV,
a) in single resolved photon processes, b) in double resolved photon
processes.
}
\end{figure}

In fig.2 we present the direct $J/\psi$ production cross sections 
through different $(c\bar c)$-pair states
 for the single resolved (fig.2a) and double
resolved (fig.2b) photon subprocesses.
We have used $m_c=1.5$ GeV for the charm quark mass   and
GRV \cite{GRV} photon distribution functions evaluated at the
factorization scale $Q^2 = \sqrt{p_T^2+4 m_c^2}$. The QCD coupling
constant $\alpha_S$ is normalized at the same scale.
We present the numerical results for the cross sections of
$1S_0$ and $^3P_J$ states production  for $p_T>1.5$ GeV in order to avoid  
the consideration of mass singularities and to deal with experimentally
observable events. 
The $J/\psi$ production cross sections in both
single and double resolved photon processes are higher than in direct
$\gamma\gamma$ collisions as can be seen when comparing fig.2 and
table I.
The color octet contribution is dominating in $J/\psi$ production in
the resolved photon case. At small values of $p_T<5$ the color octet states
$^1S_0$ and $^3P_J$ give main the contribution in direct $J/\psi$ production.
As in  hadroproduction, the  $^3S_1^{(8)}$ octet state 
becomes essential only at large values of $p_T$ ($>5$ GeV) in the
double resolved photon processes, as can be seen from fig.2b.
%
%

Note that a nondecreasing cross section of $J/\psi$ production,
 with increasing $\gamma\gamma$ invariant mass, can be obtained
 also in direct (unresolved)
photon-photon reactions when  considering higher order diagrams in pQCD with
two-gluon exchange in the $t$-channel \cite{Ginzburg}. But the 
contribution to  the total cross section of $J/\psi$ production
 from such  processes, 
$\sigma_{tot}(\gamma\gamma\to J/\psi X)\sim 2\cdot10^{-2} nb$
\cite{Ginzburg},
  is  smaller than  color octet 
contribution in single and double resolved photon subprocesses,
and decreases rapidly at large transverse momenta.

In fig.3 we present the direct $J/\psi$ production total cross
sections in dependence on the cut off value in $p_T$,  for different angular
cuts in laboratory frame ($e^+e^-$ c.m.s. system). Fig.3a corresponds 
to  $\sqrt{s_{e^+e^-}}=500$ GeV and fig.3b  to $800$ GeV.
The leading contribution in the cross sections 
with cuts $\cos{\theta}<\cos{30^\circ}$ or
 $\cos{45^\circ}$ comes from the double resolved
 photon processes, the contribution of single resolved photon
 subprocesses is negligible. 
In single resolved photon subprocesses $(c\bar c)$ octet states
 are produced  with large longitudinal momenta. 
As in the case of photon-proton
collisions, color octet states are mainly produced at the kinematical
endpoint $z=1$,  as in the case of $J/\psi$ photoproduction \cite{CK};
here $z=(p_{\gamma_2}\cdot p_{J/\psi})/(p_{\gamma_2}\cdot p_{\gamma_1})$,
where $p_{\gamma_1}$ is unresolved photon momentum and $p_{\gamma_2}$
is the momentum of the photon
which splits into a
flux of quarks and gluons. This means that the  $J/\psi$ carries 
almost the whole energy of the unresolved energetic photon and is
produced at small angles relative to the beam direction.

\begin{figure}[ht]
\vspace{-4mm}
\centering
\begin{minipage}[c]{7.5cm}
\centering
\epsfig{file=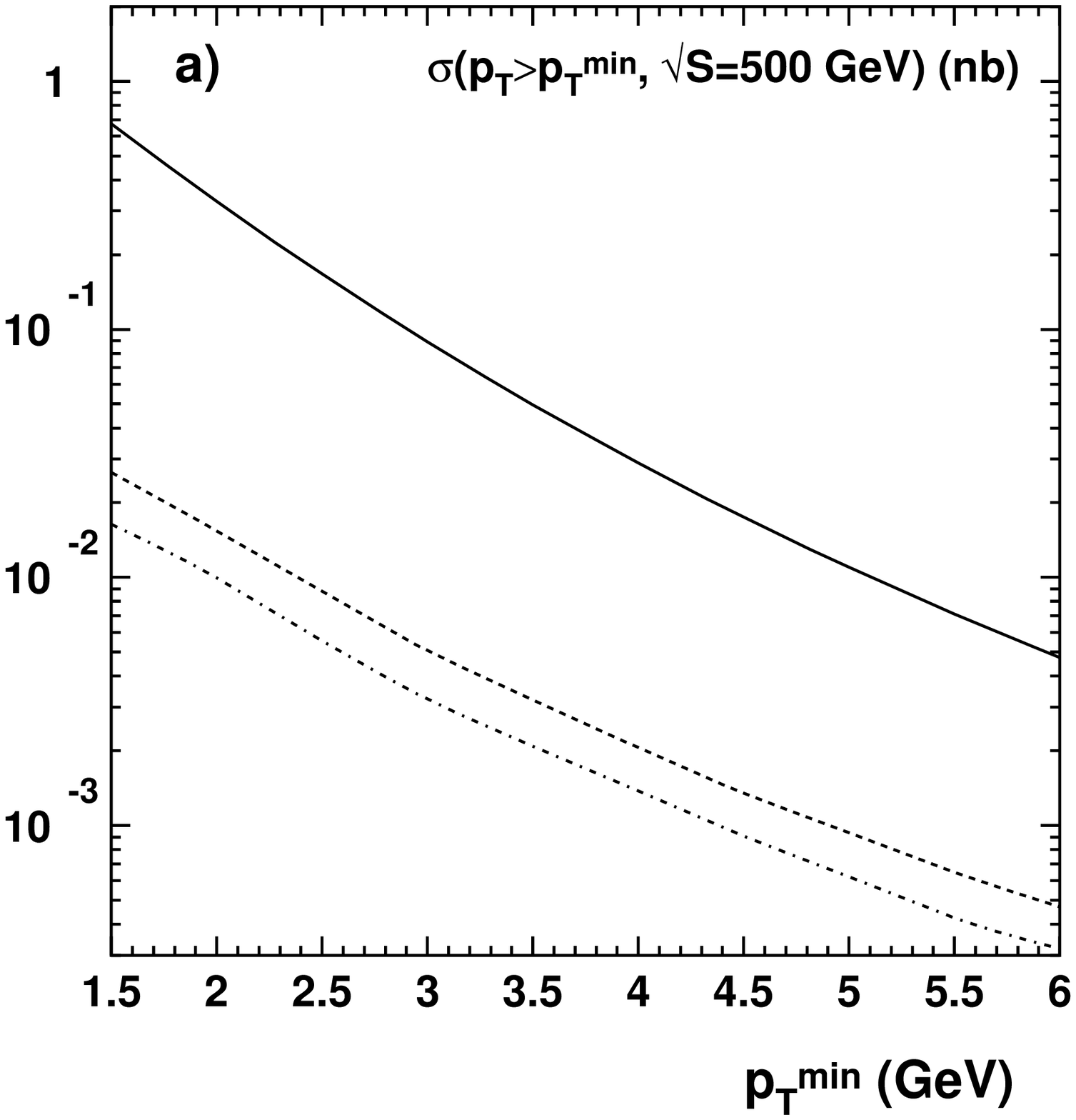,width=7.5cm}
\end{minipage}
\hspace*{0.5cm}
\begin{minipage}[c]{7.5cm}
\centering
\epsfig{file=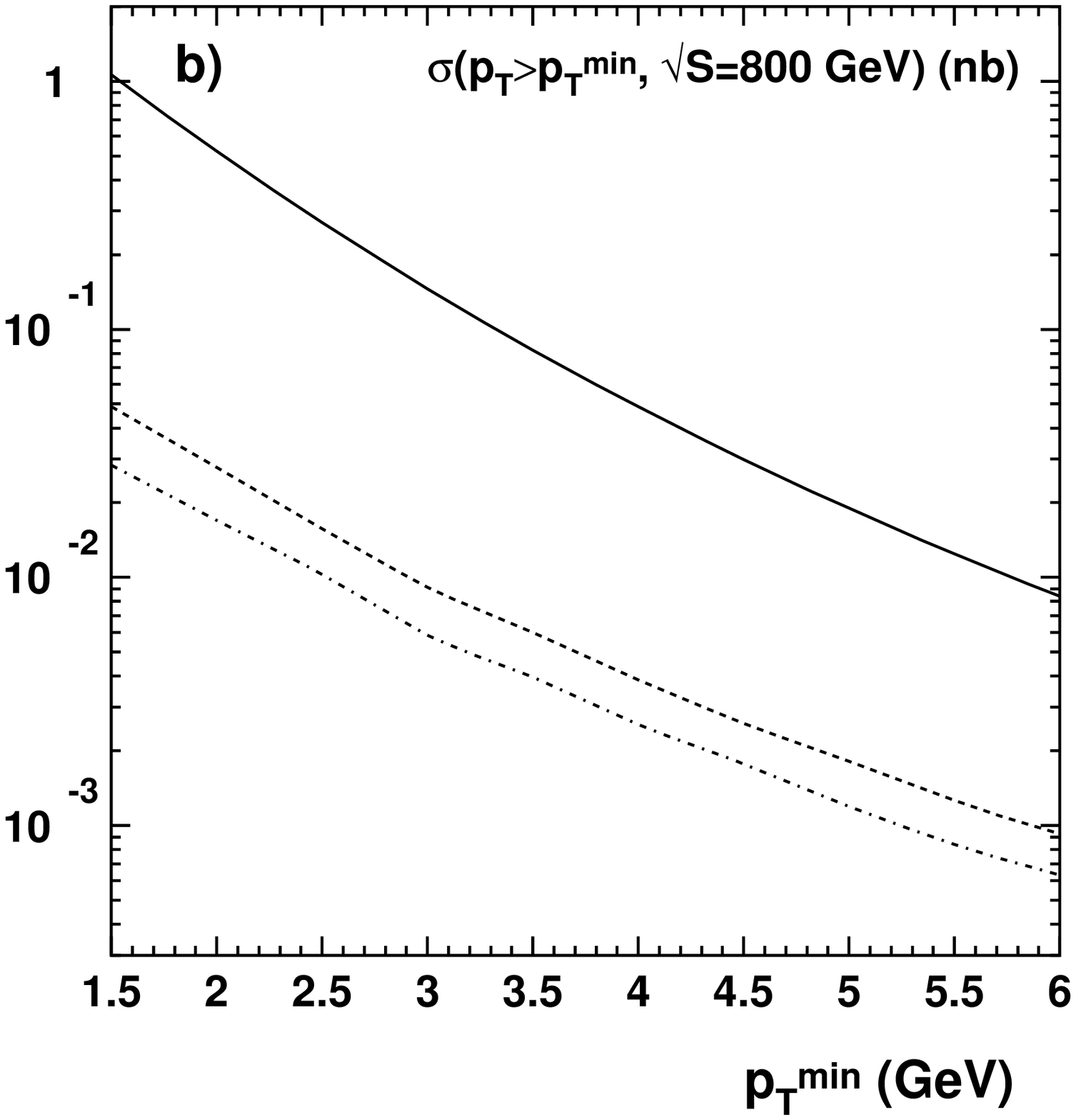,width=7.5cm}
\end{minipage}
\caption{\small
The $J/\psi$ production cross section versus minimal $p_T$ cut for
different angular cuts. The solid line corresponds to the total cross
section, the dashed line to $\cos{\theta}<\cos{30^\circ}$,
and the dashed-dotted line  to
$\cos{\theta}<\cos{45^\circ}$; a) $\sqrt {s_{e^+e^-}}=500$ GeV and
 b) $\sqrt {s_{e^+e^-}}=800$ GeV.
}
\end{figure}

The double resolved photon process is analogous to $J/\psi$
hadroproduction. At large $p_T$
the contribution from the $^3S_1^{(8)}$
color octet state dominates. Table II presents the integrated
total cross
sections and $^3S_1^{(8)}$ channel contribution for various
$p_T^{min}$ cuts. All cross sections are calculated at an angular
cut off $\cos{\theta}<\cos{45^\circ}$.
The cross sections are large enough to observe several hundred
$J/\psi$ events produced at $p_T>7$ GeV through leptonic decay channels for
 the projected integral PLC luminosity of ${\cal L}=50 fb^{-1}$.
This would allow  to extract  the long distance
matrix element $\langle{\cal O}_8^{J/\psi}(^3S_1)\rangle$ with sufficient 
accuracy \footnote{We mean only statistical significance and do not
take into account theoretical uncertainties}. 

\vspace{4mm}
{\bf Table II}. The $J/\psi$ production cross sections 
(total and through $^3S_1^{(8)}$ state), $\cos{\theta}<\cos{45^\circ}$.

\begin{center}
\begin{tabular}{|c|c|c|c|}
\hline \hline
$p_T$ & 5 GeV & 6 GeV & 7 GeV\\
\hline
total        & $6.2\cdot10^{-4} nb$ & $3.2\cdot10^{-4} nb$
 & $1.8\cdot10^{-4} nb$\\
\hline
$^3S_1^{(8)}$ & $3.8\cdot10^{-4} nb$ & $2.2\cdot10^{-4} nb$ 
& $1.4\cdot10^{-4} nb$ \\
\hline
\end{tabular}
\end{center}

Fig.4 presents the $J/\psi$ production cross section resulting from
the radiative decays of the $\chi_{cJ}$ state.
The cross sections are calculated with an angular cut 
$cos{\theta}<cos{30^\circ}$.
At  subleading order in the relative velocity expansion, $O(v^5)$, 
only one color octet state,$^3S_1^{(8)}$, contributes to the 
production of $\chi_{cJ}$ (the corresponding color octet long distance
parameter was extracted only from the Tevatron data \cite{CL}). 
The next corrections are already of order  $O(v^9)$ 
\cite{CL}. In fig.4 the color octet
(dashed-dotted line) and color singlet (dotted line)
 contributions to  $\chi_{cJ}$ meson production are presented.
\begin{wrapfigure}{l}{8.cm}
\vspace*{-7mm}
\centering
\epsfig{file=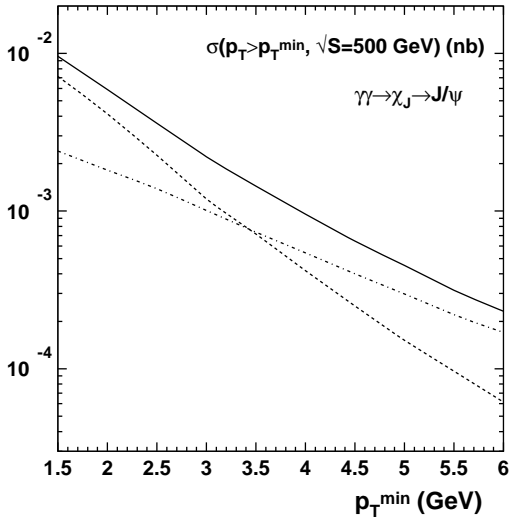,width=7.9cm}
\caption{\small
Cross section for $J/\psi$ production
resulting from radiative $\chi_{cJ}$ decay versus minimal $p_T$ cut.
 The dashed curve
corresponds to the color singlet  cross section, the dashed-dotted line
shows the $^3S_1^{(8)}$  contribution,  the solid curve
represents their sum.
}
\end{wrapfigure}
%

\section{Conclusions}
In the present paper we consider $J/\psi$ meson production at PLC. We
calculated cross sections for the production of different color octet
 and color singlet heavy
quark-antiquark states  in direct and resolved photon
processes.
In the resolved photon subprocesses already at the
 lowest order of perturbative QCD the
diagrams  arise which lead to the non decreasing
production cross sections for some quark-antiquark states with 
increasing  photon-photon c.m.s. energy.

The  contribution of color octet states to $J/\psi$ production is
dominating. It was shown that at large $p_T$ ($p_T>5$ GeV) the main
contribution to $J/\psi$ production comes from the $^3S_1^{(8)}$ state.
The  cross section is even at such values of transverse momentum
large enough to observe several hundred $J/\psi$'s through the leptonic
decay channel at a projected PLC luminocity  $50 fb^{-1}$.
This  should allow  to extract the value of the corresponding matrix
element. The cross section at smaller values of
transverse momenta can also give valuable information about the color
octet long distance matrix elements. 
The production cross section of $J/\psi$ mesons through $\chi_{cJ}$
states was also calculated. 
It is worth mentioning that background from $b$-quark decays are
smaller or at same order as direct $J/\psi$ production \cite{Eboli}
but the analysis of possible backgrounds and ways of their rejection are
subject of an additional study.

We are grateful to I.~Ginzburg, G.~Jikia, W.-D.~Nowak and V.~Serbo
for helpful discussions.
A.T. acknowledges the partly support of this work by the Alexander von
Humboldt foundation. This work was supported in part by the National
Science Foundation under Grant No. HRD9450386,
Air Force Office of Scientific Research under Grant
No. F49620-96-1-0211, and Army
Research Office under Grant No. DAAH04-95-1-0651.


\end{document}